*Review*

# A Primer on Spacetime Singularities I: Mathematical Framework

**Jean-Pierre Luminet**

Aix-Marseille Université, CNRS, Laboratoire d'Astrophysique de Marseille (LAM); jean-pierre.luminet @lam.fr

**Abstract**

This article presents a comprehensive and rigorous overview of spacetime singularities within the framework of classical General Relativity. Singularities are defined through the failure of geodesic completeness, reflecting the limits of predictability in spacetime evolution. The paper reviews the mathematical structures involved, including differentiability classes of the metric, and explores key constructions such as Geroch's and Schmidt's formulations of singular boundaries. A detailed classification of singularities—quasi-regular, non-scalar, and scalar—is proposed, based on the behavior of curvature tensors along incomplete curves. The limitations of previous approaches, including the cosmic censorship conjecture and extensions beyond General Relativity, are critically examined. The work also surveys the major singularity theorems of Penrose and Hawking, emphasizing their implications for gravitational collapse and cosmology. By focusing exclusively on the classical regime, the article lays a solid foundation for the systematic study of singular structures in relativistic spacetimes.

## 1. Definition of Singularities

*1.1. Splendour and Misery of Singularities*

The problem of singularities in four-dimensional space-time is undoubtedly one of the most intriguing in General Relativity, especially since the numerous "twists" of the last decades on the subject.

While the concept of singularities is well established in classical field theories (hydrodynamics, electrodynamics, etc.), the same cannot be said of General Relativity. The difference lies in the following fact: in all other field theories, space-time is given once and for all as the framework for the theory, and a singularity in the field generally only indicates a limitation of the theory; for example, in electrodynamics, a solution of Maxwell's equations is said to be singular if the field is infinite, and therefore indeterminate, at some point in Minkowski space-time. By analogy, we would like to be able to say in General Relativity that space-time is singular if the metric tensor is indeterminate at a point.

The trouble is that General Relativity is a theory in which the field interacts with the framework in which it is described: the field itself determines the geometric properties of space-time.

In fact, there is always the possibility of escaping the problem, and many physicists do not believe that Physics can collapse at singularities. Various more or less viable hypotheses have been put forward to this effect:

1) General Relativity does not predict singularities.

This view was widely held in the 1960s (Lifshitz and Khalatnikov, 1963). However, the study of gravitational collapse and the publication by Penrose (1965) and Hawking

(1966) of theorems demonstrating that in certain circumstances a space-time singularity necessarily occurs (if the theory of General Relativity is valid and if the momentum-energy tensor of matter satisfies a positivity condition) provided a striking counter-example to Lifshitz and Khalatnikov's conjecture. The latter courageously admitted their error and became among the most fervent supporters of the existence of singularities in cosmology (Lifshitz and Khalatnikov, 1970).

*2)* General Relativity can also be modified to avoid singularities. The necessary modifications must include the possibility of repulsive gravity (since it is the attractive character of gravity that seems to be at the origin of the existence of singularities). In the theory of Brans and Dicke (1961), gravity remains attractive and singularities are predicted (Hawking and Penrose, 1970). The Einstein-Cartan theory (see for example Hehl et al., 1976) contains a spin-spin interaction that can be repulsive; however, there are still situations - such as purely gravitational and electromagnetic fields - in which singularities occur. Most other theories of gravitation appear in some way to conflict with observations or involve assumptions that are physically unacceptable.

3) The cosmic censorship hypothesis: after Blaise Pascal, who postulated that Nature abhors vacuum, this conjecture puts forward the idea that Nature abhors 'naked' singularities. In other words, if we start from an initially non-singular and asymptotically flat situation, all the singularities likely to develop later as a result of gravitational collapse would be hidden from the view of an observer placed at infinity behind a horizon (Penrose, 1969). Under these conditions, the singularities could rightly be ignored, since they would cause no detectable effect for observers careful enough not to fall into a black hole.

The cosmic censorship hypothesis has been beaten back on several fronts. Ethically speaking, it corresponds to a perfectly selfish attitude towards unlucky observers who might actually fall into a black hole. Physically, it does not solve the problem of the initial singularity of the universe ("big bang"), which is naked. Finally, Hawking's discovery (1974, 1975) of the process of particle creation by a black hole cast serious doubt on the validity of this hypothesis. Studies on strong cosmic censorship examine whether singularities remain hidden or become observable. Works by Christodoulou (1984, 1999), Dafermos (2014), and others – for recent reviews, Landsman (2021), Van de Moortel (2025) – have revealed complex structures.

4) General Relativity could eventually be quantified.

When this vast and difficult program was undertaken, some physicists hoped that strong quantum effects could eliminate singularities. Hawking (1976) argued that space-time singularities represent a fundamental limitation on our ability to predict the future - a limitation analogous to that imposed by the uncertainty principle in Quantum Mechanics. However considerable theoretical work has being done on how quantum gravitational effects might modify the nature of singularities, or even eliminate them (for a recent review, see e.g. Crowther and De Haro, 2022).

In the present work we will not tackle the problem of the quantization of General Relativity. All the results will be formulated in a classical (non-quantum) framework. In this simpler context, problems relating to the occurrence and description of singularities have been addressed for the last decades. But while positive results were quickly found for their occurrence (Penrose and Hawking, already cited), the road to a good description of singularities was strewn with pitfalls: no sooner had a sketch of a description been formulated than counter-examples flourished in the field of publications.

Finally, after much trial and error, concepts were transformed into definitions, conjectures into theorems, arguments into proofs and hopes into results; in short, beliefs into knowledge...

In the following, we will attempt to untangle the skein, focusing only on the mathematical background without discussing the internal structure of singularities or the behaviour of spacetime models near singularities, a subject we will cover in a second part of this review article. As most of the mathematical basis for dealing with classical spacetime singularities was provided in the 1970's, most of the references quoted in this paper date from this period. General review work on classical space-time singularities without mathematical details but more recent references can be found e.g. in Joshi (2014) and Curiel (2019).

*1.2. The Notion of Singularity*

Before thinking about giving a good definition of what a singularity is in General Relativity, we need to have a notion of it.

A singularity can always be considered as the violation of a regularity condition of a basic structure of General Relativity. The structures in question can be the metric structure, the conformal structure and the affine structure.

Given a model (M,g) and initial conditions on a compact spatial surface, we say that we obtain a singular development if (M,g) satisfies at least one of the following conditions:

i) (M,g) contains strong discontinuities (violation of the junction conditions)

ii) (M,g) is acausal (there are closed timelike lines)

iii) (M,g) is incomplete (the history of some particles cannot be described for all values of proper time).

For each of these structures, there is a complicated hierarchy of regularity conditions.

This notion of singularity is negative in the sense that it does not tell us what singular behaviour actually is. This is precisely what we shall try to clarify through rigorous mathematics.

Note that case i) is a purely local property, case ii) is essentially global, and case iii) is in principle global, but incompleteness can be "localized" by creating an extension of space-time in which all incomplete curves have extremities and by attaching a singularity to an equivalence class of incomplete curves (Schmidt, 1971).

In commonly accepted definitions of singularities, case (ii) is omitted: for good reasons, the problem of causality is treated in a different context from that of singularities (see for example Hawking and Ellis, 1973, §6), and we will not discuss it here. The emphasis will be placed on the incompleteness of space-time.

This leaves the equivalence: a singularity is an incompleteness of space-time whose metric satisfies certain regularity conditions.

Let us briefly explain what these conditions are.

*1.3. Discontinuities in the Metric Structure*

We may assume the manifold M of class $C^\infty$, since, according to Whitney's embedding theorem, every atlas of class $C^k$ (k≥1) contains a sub-atlas of class $C^\infty$. But the problem of choosing the $C^\infty$ structure is not so simple; although $C^\infty$ structure denotes a smooth atlas, a wrong choice of a particular coordinate charts can lead to factice "coordinate singularities".

Concepts as diverse as geodesics, the Cauchy problem or the continuity of the stress-energy tensor require, in order to be well defined, certain regularity conditions which will be expressed by *classes of differentiability* of the metric g.

Recall that a function f defined on an open O of $R^n$ is said to be *locally Lipschitzian* if for any open $U \subset O$ of compact closure, there exists a constant K such that $\forall\ p,q \subset U : |f(p) -f(q)| \leq K\ |p - q|$, where |p| denotes the Euclidean norm $\{\Sigma(x^i(p)^2\}^{1/2}$. A morphism $\phi$ is said to be locally Lipschitzian and denoted $C^{1-}$ if the coordinates of $\phi(p)$ are locally Lipschitzian functions of the coordinates of p. Similarly, we say that ϕ is $C^{r-}$ if it is $C^{r-1}$ and if the derivatives of order r-1 of the coordinates of ϕ(p) are locally Lipschitzian functions of the coordinates of p.

$C^0$ means continuous, $C^{0-}$ means locally bounded.

A spacetime (M, g) is said to be of class *$C^k$ (resp. $C^{k-}$)*, k≥0, if the $C^\infty$-manifold M is Hausdorff and if:

i) the components of the metric $g_{ab}$ are continuous and admit locally bounded derivatives

(ii) the components of the curvature tensor $R^a{}_{bcd}$ are functions of class *$C^k$ (resp. $C^{k-}$).*

To guarantee the existence and uniqueness of the geodesics, the following assumption is made:

$g_{ab}$ is of *strong curvature differentiability class sc-$C^{2-}$*, i.e. there exists an atlas $C^\infty$ of M such that the metric tensor is Lipschitzian (*$C^{1-}$*) and the Riemann tensor $R^a{}_{bcd}$ is locally bounded (*$C^{0-}$*) [Clarke, 1975b].

In fact, a sc-$C^{2-}$ space-time can be considered as the limit of a sequence of $C^{2-}$ space-times. Any result on geodesics that can be formulated in terms of the Riemann tensor alone will be valid on a sc-$C^{2-}$ spacetime as a limiting case of a sequence of geodesics.

To prove the existence and uniqueness of the solution of the Cauchy initial data problem, we use the hypothesis:

There exists an atlas $C^\infty$ of M such that g and its derivatives up to order four exist in the sense of distributions and are locally square integrable; g is then said to be of *Sobolev class $W^4$* (Sobolev, 1963; see also Hawking et Ellis, 1973, §7).

In order to obtain a dependence of class $C^k$ in the initial data, these conditions must be strengthened in sc-$C^{2+k-}$ and $W^{4+k}$.

In order to ensure the continuity of the normal components of the stress-energy tensor $T^a{}_b f_{,a}$ through the {f = const} surfaces, we must impose *junction conditions*: g is of class $C^1$, $C^3$ by pieces (see for example Misner, Thorne et Wheeler, 1974, pp. 551-556).

Finally, it should be noted that all these assumptions must be slightly weakened when describing certain phenomena such as shock waves (Choquet-Bruhat, 1968).

In the following sections, it will always be assumed that the metric has the required differentiability class. In general, we will use $C^0$-space-times (resp.$C^{0-}$) when there are coordinates in which the $g_{ab}$ are functions of class $C^2$ (resp.$C^{2-}$), corresponding to continuous (resp. locally bounded) Riemann tensors. The second case admits shock waves, the first does not.

The classification of singularities that we will describe later is designed to take account of both discontinuities in the metric and irregularities in the curvature tensor.

*1.4. Space-Time Completeness*

We have already seen that in General Relativity, the difficulty arises from the fact that we want to describe the singularities of a field which itself determines the geometric properties of space-time.

The fact that components of the metric tensor become infinite at a point may simply be due to a poor choice of coordinate system. In any case, we can always amputate the manifold of a region in which the metric tensor is not defined and say that the remaining manifold represents the whole, non-singular spacetime. This shows that a definition of a singular spacetime cannot be based on the behavior of tensor field components and that singular points cannot be represented as belonging to the spacetime manifold itself. Since, on the other hand, we can always perform amputation surgery on space-time, we would like to be sure that in such an operation no non-singular part has been removed.

Recognizing whether or not regions have been cut off from space-time relates precisely to the problem of the completeness of space-time.

*1.5. Riemannian Manifolds*

The case of a manifold M with a Riemannian metric g (positive definite) is very simple. A *distance function* d(p,q) can be defined as the lower bound of the lengths of the curves from p to q:  $d(p,q) = \inf(\int_{pq} ds)$.

This function defines a metric in the topological sense, i.e. the balls B(p,r) = {q ∈ M | d(p,q)<r} form a basis of open sets for the topology of M.

A *Cauchy sequence* on M is an infinite sequence of points $p_n \in M$ such that $\forall \varepsilon > 0$, $\exists N | \forall n, m > N$, $d(p_n, p_m) < \varepsilon$.

We say that (M,g) is *m-complete* (*m* for "metric") if every Cauchy sequence with respect to the distance function d converges to a point of M.

An equivalent formulation is as follows: (M,g) is m-complete if any curve $\gamma$ : [a,b] →M of class C[1] and finite length has an extremum, i.e. there exists a point p of $\gamma$ such that for any neighborhood V of p, there exists t ∈ [a,b] such that $\forall t_1 \in [t,b]$, then $y(t_1) \in \gamma$.

Singularities on a Riemannian manifold can therefore be described by the Cauchy completion, which exists and is unique.

We can also define a geodesic completeness (g-completeness) on a Riemannian manifold: (M,g) is g-complete if any geodesic of M can be extended to arbitrarily large values of its affine parameter.

Recall that a curve $\gamma(t)$ is geodesic if $\frac{D}{\partial t}(\frac{\partial}{\partial t})_\gamma$ is parallel to the tangent vector to the curve $(\frac{\partial}{\partial t})_\gamma$ [D/∂t is the covariant derivative along $\gamma(t)$] and that the parameter t is said to be *affine* if $\frac{D}{\partial t}\left(\frac{\partial}{\partial t}\right) = 0$.

An affine parameter on a geodesic curve is determined to within one additive constant and one multiplicative constant, i.e. within one affine transformation; in other words, if t is an affine parameter of $\gamma$, t' = at+b is an affine parameter of $\gamma$, where a and b are arbitrary constants.

It can be shown (Kobayashi and Nomizu, 1963) that for a positive definite metric, m-completeness and g-completeness are equivalent.

*1.6. Pseudo-Riemannian Manifolds*

The case of pseudo-Riemannian varieties is more complex.

A Lorentzian metric, for instance, does not define a metric in the topological sense, since there are inextensible curves of zero length (light rays). Even in the most regular case, the Minkowski space-time, inextensible curves of finite length can be found: the curve x-t = (x+t)$^{-2}$ from (x=1, t=0) to infinity has length √8.

So, m-completeness cannot be defined for pseudo-Riemannian manifolds.

On the other hand, there are always geodesics and we can obviously talk about g-completeness. However, there are three kinds of geodesics: timelike geodesics, null geodesics and spacelike geodesics, so that we are led to consider the three corresponding kinds of g-completeness:
- t-g-completeness (completeness of timelike geodesics)
- n-g-completeness (completeness of null geodesics)
- s-g-completeness (completeness of spacelike geodesics)

One might expect a priori that a spacetime g-complete in one kind is g-complete in the other two kinds. Unfortunately, this is not the case (Kundt, 1963; Geroch, 1968a). Examples have been given of spacetimes that are:
1) t-g-complete but n,s-g-incomplete
2) s-g-complete but t,n-g-incomplete

3) n-g-complete but t,s-g-incomplete
4) t,n-g-complete but s-g-incomplete
5) s,n-g-complete but t-g-incomplete

To my knowledge, no example has yet been given of a t,s-g-complete but n-incomplete spacetime, and the non-existence of such a spacetime has not been demonstrated.

*1.7. Physical Significance of Geodesic Incompleteness*

The incompleteness of timelike geodesics has an immediate physical meaning: there are particles or free observers whose worldlines no longer exist after (or before) a finite interval of proper time. It seems appropriate to consider a spacetime containing such geodesics as singular.

Similarly, null geodesics are the trajectories of particles with zero rest mass. On the other hand, since the physical existence of free "tachyons" moving on spacelike geodesics has not been proven, the incompleteness of spacelike geodesics has no physical meaning for the moment.

It is therefore tempting to adopt the view that the completeness of timelike and null geodesics is the minimal condition for a spacetime to be non-singular.

Considering the incompleteness of timelike or null geodesics as indicating the presence of singularities has the advantage of leading to a number of fundamental theorems relating to the occurrence of this kind of singularities. These theorems, due to Penrose (1965), Hawking (1966, 1967) and Geroch (1966) gave rise to a number of important works on the description of predicted singularities.

Hawking (1966) and then Geroch (1968b) attempted to structure the set of singularities (i.e., relate it to the spacetime manifold, then provide it with a topology and possibly a metric) using the notion of geodesic incompleteness.

Note, however, that the problem of attaching a boundary to the space-time (M, g) so as to describe singularities only concerns boundary points located at *finite* distance from points of M. This boundary will not therefore represent points "at infinity" such as those of the conformal boundary studied by Geroch, Kronheimer and Penrose (1972).

*1.8. Geroch's Construction*

This consists in giving a structure to the set of incomplete geodesics. Roughly speaking, these incomplete geodesics are arranged in equivalence classes, each class defining a point in an abstract space ∂. The basic idea is that an incomplete geodesic γ can be defined by its initial conditions: any point on γ and the tangent vector to γ at that point. Let's consider the set of geodesics resulting from small variations in these initial conditions. This bundle is called the *thickening* of γ (fig. 1).

The two incomplete geodesics $γ_1$ and $γ_2$ are said to belong to the same equivalence class (and thus define the same singular point of ∂) if any thickening of one intersects any thickening of the other.

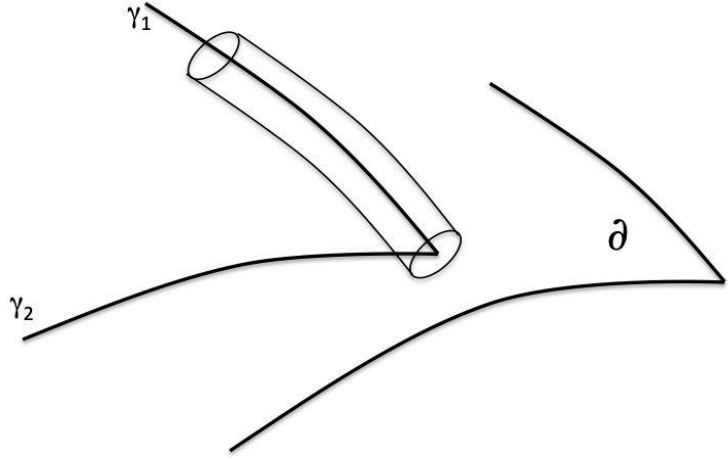

**Figure 1 : Thickening of incomplete geodesic γ**

Let us now explain Geroch's method more precisely. Let M be a spacetime. A geodesic in M is determined in a unique way by giving a point p of M and a non-zero vector $X_P \in T_P(M)$: the geodesic starts from p with an initial direction $X_P$ and an affine parameter $\lambda$ determined by the equation:

$dx^i/d\lambda |_p = X_P^i$ (i= 0,1,2,3)

We denote G the set of pairs (p, $X_P$); G is the part of the tangent fiber bundle of M comprising only the non-zero vectors; we therefore call it the *reduced tangent fiber bundle* of M. There is thus a one-to-one correspondence between the elements of G and the geodesics of M. These geodesics have an extremum, they are parametrized by an affine parameter, the latter cancelling at the extremum and being positive elsewhere along the curve.

We now want to characterize the subset $G_I$ of G corresponding to the incomplete geodesics of M.

To do this, we define the scalar field Φ on the manifold G to be the total affine length of the corresponding geodesic in M. $G_I$ is therefore the subset of G for which Φ is finite.

With a view to constructing a topology on $G_I$, we introduce the manifold H = G×[0, ∞] and the following two subsets of H:

$H_+ = \{(p, X_P, a) \in H \mid \Phi(p, X_P) > a\}$
$H_0 = \{(p, X_P, a) \in H \mid \Phi(p, X_P) = a\}$

Note that in particular the points (p, $X_P$, Φ(p,$X_P$)) are in $H_0$.

There is a natural morphism Ψ : $H_+$ →M which maps to a point (p, $X_P$, a) in $H_+$ the point in M constructed from p by measuring the affine length a along the geodesic (p, $X_P$).

This application will allow us to define the openings of a topology on $G_I$. Let O be an open of M. Consider the subset S(O) of $G_I$ defined by:

S(O) = {(p, $X_P$)∈ $G_I$ |there exists an open U in H containing the point (p, $X_P$, Φ(p,$X_P$)) of $H_0$ such that Ψ(U $H_+$)⊂O}

We can check that given two open sets $O_1$ and $O_2$ of M, we have S($O_1$) ∩ S($O_2$) = S($O_1$ ∩ $O_2$). So, when O crosses the set of open sets of H, S(O) is a basis of open sets of a $G_I$ topology.

This topology is used to form equivalence classes between elements of $G_I$ : if α and β are two elements of $G_I$, α~β if any open set of $G_I$ containing α contains β**,** and vice versa. This is an equivalence relation. The set of equivalence classes is denoted ∂ and is called *g-boundary* (g for "geodesic" or for "Geroch", take your pick!).

For example, the elements (p, $X_P$) and (p, $\lambda X_P$) of $G_I$, where $\lambda$ is a non-zero constant, belong to the same equivalence class.

The topology defined on $G_I$ induces a topology on $\partial$ (the quotient-topology). Note, however, that there are various ways of forming equivalence classes to form a g-boundary. The identification described above is the "weakest" in the sense that the points of $G_I$ are identified only if they are topologically indistinguishable, i.e. if they always appear in the same openings; the resulting g-boundary satisfies the separation axiom $T_0$ (given any two distinct points x and y of $\partial$ there exists either a neighborhood of x that does not contain y or a neighborhood of y that does not contain x). Stronger identifications could define g- boundaries satisfying the separation axioms $T_1$ (given any two distinct points x and y of $\partial$, there exists a neighborhood of x that does not contain y and a neighborhood of y that does not contain x), $T_2$ or Hausdorff (there exists a neighborhood of x and a neighborhood of y that are disjoint), or $T_3$ ($\partial$ satisfies axiom $T_1$ and is regular, i.e. for any point x of $\partial$ the closed neighborhoods of x form a basis of neighborhoods of x).

We now attach $\partial$ to the space-time M by defining the disjoint union $\bar{M} = M \cup \partial$. The opens of $\bar{M}$ are the subsets (O,U) where O is an open of M and U is an open of $\partial$ such that $U \subset S(O)$.

We check that the intersection of two openings of $\bar{M}$ is an open set of $\bar{M}$; we therefore have a basis of open sets for a topology on $\bar{M}$. Space-time with its g-boundary $\bar{M}$ is a manifold with an edge.

According to Geroch, the g-boundary consists of the "singular" points of space-time M. A proposition such as "the mass density becomes infinite at the singularity" translates mathematically as: "the point x of the g-boundary has the property that whatever the number $\rho_0$, there exists an open neighborhood (O,U) of x in $\bar{M}$ such that $\rho > \rho_0$ in O",

Geroch then defines new structures on the g-boundary. Indeed, the topological properties of $\partial$ are not sufficient to classify singularities. Geroch therefore begins by giving $\partial$ a causal structure, i.e. determining on the one hand the "past" and "future" of each point of $\partial$, and defining on the other hand the "spatial" g-boundaries and the "temporal" g-boundaries.

Finally, Geroch provides $\partial$ with a differentiable structure and a metric, subject to certain conditions on space-time.

*1.9. The Sad Story of a Rocket*

We will not go into the details of these constructions here. Indeed, Geroch's attempt, interesting as it is, has certain limitations that the author himself acknowledges at the end of his article. We have already mentioned that there are several ways of defining equivalence classes on $G_1$, hence a certain ambiguity in the local description of singular points. But the major obstacle is that the class of geodesically incomplete spacetimes does not cover all the kinds of singularities we would like to see.

Let's imagine an inextensible non-geodesic time curve with bounded acceleration and finite proper length. Could we not describe as singular the situation of the passenger in a rocket who, having travelled the entire length of the curve in a finite proper time, would no longer be represented by a point in the space-time manifold? Geroch himself constructed a geodesically complete space-time containing such a curve (Geroch, 1968a).

We would like to be able to say that this space-time is singular. So, a definition of the concept of singularity must include all incomplete curves, geodesic or not.

The most natural idea is to consider a notion of completeness according to which all curves of class $C^1$, of finite length measured by an appropriate parameter, have an end.

*1.10. Generalized Affine Parameter*

The parameter in question used to measure the length of curves must be a generalization of the concept of affine parameter, which until now has only been defined for geodesics. The concept of a generalized affine parameter is therefore introduced as follows:

Let $\gamma(t): [a,b] \to M$ be a curve of class $C^1$ passing through a point p of M and let $\{E_i\}_{i=0,1,2,3}$ be a basis of the tangent space at p, $T_P(M)$.

The basis $\{E_i\}$ can be shifted parallel to itself along $\gamma(t)$ to obtain a basis of $T_{\gamma(t)}$ for all t in [a,b].

The tangent vector to the curve, $V = (\partial/\partial t)_{\gamma(t)}$ can therefore be expressed in terms of this basis as $V = V^i(t)E_i$.

We then define on $\gamma(t)$ a generalized affine parameter v by:

$v = \int_P ||V(t)|| dt$,

where $||V(t)||$ denotes the Euclidean norm of the tangent vector V:

$||V(t)|| = \{\Sigma_i (V^i(t))^2\}^{1/2}$.

If $\gamma$ is a geodesic, v reduces to an affine parameter.

We can see that v depends on the point p and the basis $\{E_i\}$. However, it is easy to check that if v' is another generalized affine parameter defined from a basis $\{E_i'\}$ in p, the length of the curve g is finite in the parameter v' if and only if it is finite in the parameter v. We therefore do not restrict the problem by considering only orthonormal bases $\{E_i\}$.

*1.11. b-Completeness*

We can now define b-completeness (b for "bundle"; see §1.12):

(M, g) is b-complete if any curve of class $C^1$ of finite length measured by a generalized affine parameter has an extremum.

Note that if g is positive definite, the generalized affine parameter reduces to the curvilinear arc length, so that b-completeness, g-completeness and m-completeness are all equivalent.

For a Lorentzian metric, b-completeness obviously entails g-completeness, the opposite being false.

The completeness of spatial curves (s-completeness) leads to the completeness of null curves (n-completeness) and time curves (t-completeness). Thus, the affine completeness of all curves is equivalent to the completeness of spatial curves alone.

On the other hand, s-g-completeness does not imply t-g- or n-g-completeness.

The completeness of time curves leads to the completeness of curves with bounded acceleration (b.a-completeness), which in turn leads to the completeness of time geodesics. Note also that it is logical to call the completeness of null and timelike geodesics (t,n-g-completeness) "the completeness of causal geodesics" (c-g-completeness), for obvious physical reasons. The Hawking and Penrose singularity theorems involve incomplete causal geodesics.

Finally, it can be shown that the causal completeness of all curves is in fact equivalent to the completeness of null curves (see, for example, Seifert, 1975).

We can thus schematize the hierarchy between the various types of completeness as follows (the ~ sign stands for "completeness"; arrows deduced by transitivity are not indicated), see Fig. 2:

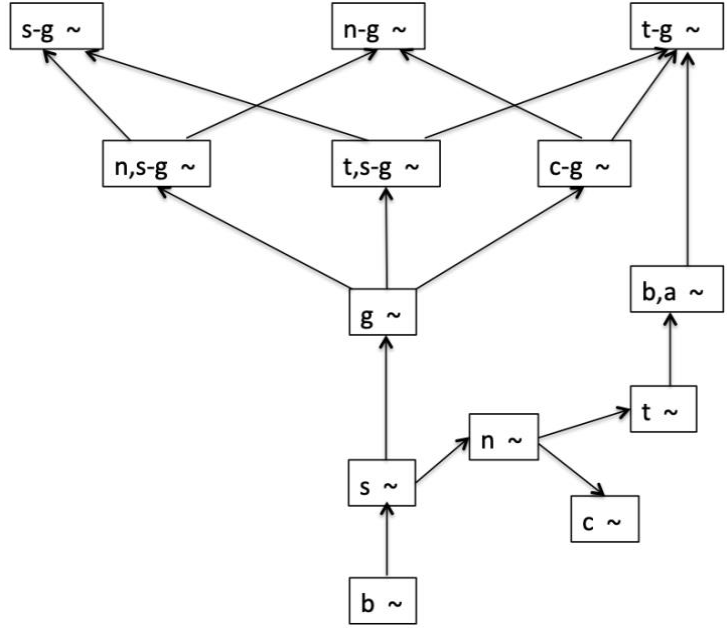

**Figure 2. Hierarchy between the various types of geodesic completeness**

*1.12. Schmidt's Construction.*

The rigorous construction of a b boundary of space-time representing the set of extremities of all incomplete curves of M was carried out by Schmidt (1971, 1972). His very elegant and sophisticated method is based on the fibred structure of linear reference frames L(M).

Recall that a fibred structure (see for example Steenrod, 1951 and Hicks, 1965) on a topological space E (total space or fibred) is given by:

a) a topological space B called base and a topological space F called fiber bundle-type

b) a continuous surjective application p: $\pi : E \rightarrow B$ called *projection*. For $p \in B$, $\pi^{-1}(p)$ is called *fiber of p.*

c) a covering of the basis B by a family of open U such that $\pi^{-1}(U)$ is homeomorphic to U×F, i.e. for any point p in U, there exists a homeomorphism $\Phi_P : \pi^{-1}(p) \rightarrow V$ such that the application $\psi : \pi^{-1}(U) \rightarrow U \times S$ is defined by $\psi(x) = (\pi(x), \Phi_{\pi(x)})$

Thus, the fibered structure generalizes the notion of topological product. Indeed, the topological product E = X×Y is a fibred bundle with basis X, type-fiber Y, projection π(x,y) = x ; the covering reduces to the single open U = X, the homeomorphism ψ is the identity, the structural group G reduces to the identity.

An application f : B→E such that π∘f is the identity function on B is called a section of the fibered E. A fibered space where G = F is called a *principal fibered bundle.*

Consider a differentiable manifold $V_n$. A linear reference frame $R_p$ at a point p of $V_n$ is an ordered basis $E_1,…,E_n$ of the tangent space $T_p(V_n)$. The set $L(V_n)$ of all linear reference points at all points of $V_n$ can be given a differentiable fibered structure: the basis is $V_n$, the type-fiber is the linear group of order n GL(n,R) (isomorphic to the group of regular n×n matrices with real coefficients), the projection is defined by $\pi(p,R_p) = p$, and the structural group is GL(n,R).

$L(V_n)$ is therefore a differentiable main fibred bundle called *fibered bundle of linear reference frames on manifold* $V_n$. The structural group GL(n,R) acts on the right-hand side as follows:

$A = \{(a^i_j)\} \in GL(n,R)$ transforms $(p,R_i)$ into $(p, A_{ij}E_j)$.

We can define a subbundle of L(V$_n$), the *fiber of orthonormal reference frames O(V$_n$)*. In the case of a Lorentzian metric, the structural group is then the Lorentz group of dimension n: O(n-1,1).

If L(V$_n$) admits a section, V$_n$ is said to be parallelizable. It can be shown that the existence of a connection on V$_n$ implies that the manifold V$_n$ is parallelizable.

Let us now go through the various stages of Schmidt's construction, without however dwelling too much on the very technical aspects which have been amply described in the literature (Hawking and Ellis, 1973 chap.8).

Consider a space-time manifold M and its orthonormal reference frame O(M). Since M has a connection, we can define so-called *horizontal* curves in O(M), obtained by transporting an orthonormal frame of reference in parallel along the curves of M.

Let T$_u$(O(M)) then be the tangent space at the point u of O(M), of dimension 10. The vectors tangent at u to the horizontal curves of O(M) passing through u give rise to a vector subspace of T$_u$(O(M)), of dimension 4, called the *horizontal subspace H$_u$(O(M))*.

T$_u$(O(M)) then decomposes into the direct sum of the horizontal subspace *H$_u$(O(M))* and the *vertical subspace V$_u$(O(M))* consisting of the vectors of T$_u$ that are tangent to the fiber $\pi^{-1}(\pi(u))$.

This makes it possible to define a canonical set of linearly independent vectors constituting a basis of T$_u$(O(M)) ; such a basis $\{G_A\}_{A=1,2,...,10}$ can be written $\{G_A\} = \{E_A, \tilde{E}_i\}$ where $\{E_a\}_{a=1,...,4}$ is a basis of H$_u$ and $\{\tilde{E}_i\}_{i=1,...,6}$ is a basis of V$_u$.

With this basis, we can define a positive definite metric e on O(M) by :

e(X,Y) = $\Sigma_A$ X$^A$Y$^A$, where X,Y $\in$ T$_u$(O(M)} and X$^A$, Y$^A$ are the components of X,Y in the basis $\{G_A\}$ .

This metric makes it possible to introduce a distance function d on O(M): u and v being two points of O(M), d(u,v) is the lower bound of the lengths (measured by e) of the curves joining u and v.

O(M) thus becomes a metric space.

As we saw in §1.5, we can then complete O(H) by means of Cauchy sequences, defining the Cauchy completion $\overline{O(M)}$ as the set of equivalence classes of Cauchy sequences in O(M). $\overline{O(M)}$ is unique to within one isomorphism. We can write $\overline{O(M)}$ = O(M)$\cup\partial$O(M), where $\partial$O(M) = {u $\in$ $\overline{O(M)}$ | u $\notin$ O(M)}.

The action of the Lorentz structural group O(3,1) can be extended to $\overline{O(M)}$. We call $\bar{M}$ the set of orbit $\overline{O(M)}$ under the action of O(3,1), and $\partial$M the set of orbits of $\partial$O(M). The projection p $\pi$ : O(M)$\rightarrow$M can be extended to $\bar{v}: \overline{O(M)} \rightarrow \bar{M}$ by defining $\bar{\pi}$(u) as the orbit passing by u $\in$ $\overline{O(M)}$.

$\overline{O(M)}$ being provided with a topology induced by the metric e, $\bar{M}$ can also be provided with a topology by defining an open U of such that $\pi^{-1}$(u) is an open set of $\overline{O(M)}$. This is the weakest topology for which p is continuous.

We thus have a method for attaching a boundary $\partial$M to the space-time M.

However, it would be premature to call all the points of $\partial$M "singular points". Indeed, M could, for example, be extensible, so that singular points of $\partial$M could be perfectly regular in an extension of M.

So Schmidt was led to define a singularity as a point of $\partial$M that is contained in the boundary $\partial\tilde{M}$ of any extension $\tilde{M}$ of M.

The fundamental question is to prove that $\partial$M coincides with the b-boundary consisting of the extremities of all incomplete curves of M. We have the theorem (Schmidt, 1971):

(O(M), e) is incomplete if and only if (M,g) is b-complete.

The proof of this theorem depends on the fact that every spacetime M is locally b-complete, *i.e.* every point of M has a neighborhood U of compact closure $\bar{U}$, such that the b-boundary ∂U of U coincides with the boundary $\bar{U}$/U of U in M (Schmidt,1972).

We will not go into the details of the proofs here, as these can be found in the classical literature already mentioned.

Thus, any equivalence class of b-incomplete curves in M defines a point of ∂M.

Note that the topology of $\bar{M}$ is not necessarily Hausdorff, *i.e.* there may exist a point p ∈ M and a point q ∈ ∂M such that any neighborhood of p in M intersects any neighborhood of q. This situation occurs when the point q corresponds to an incomplete curve which is totally or partially trapped in M, as happens for example for the Taub-NUT space (for more details, see §2.4 and Hawking-Ellis, 1973, §8.5).

Other properties of the b-boundary have recently been discussed (Friedrich, 1974).

## 2. Classification of Singularities

*2.1. Introduction*

There is a feeling that the somewhat abstract definition of singularities in Schmidt's sense must now be related to the intuitive physical conception one may have of a singularity, namely a point where the gravitational forces due to curvature become infinite.

We must therefore distinguish, among the points of the b-boundary, those where the Riemann tensor is irregular and thus establish a *classification of singularities* (Clarke, 1973, 1975a; Ellis and Schmidt, 1977).

The idea is to base the classification on the behaviour of the components of the Riemann tensor in a tetrad transported along an incomplete curve defining the singularity.

*2.2. Extensible Manifolds*

Before undertaking a classification of the points of the b-boundary, we must ensure that the space-time manifold (M, g) is indeed *inextensible* (or *maximal*).

As already mentioned in §1.12, a point p on the b-boundary ∂M of an extensible spacetime M may be such that there exists an isometry $\psi$ of (M, g) in a larger spacetime (M', g') which sends p onto an interior point of M': $\bar{\psi}(p) \in M'$ (where $\bar{\psi}$ is the extension of $\psi$ on ∂M). Clearly, there is nothing "singular" about such a point in the physical sense. Clarke (1975b) named such points "locally inextensible inessential singularities", but it is clear that the term "singularity" is superfluous. We will simply call such a point a *regular boundary point*.

Let us give two typical examples of such pathologies:

(a)- The Minkowski half-spacetime

The metric is: $ds^2 = -dt^2 + dx^2 + dy^2 + dz^2$, defined for $-\infty < x,y,z < +\infty$ and $0 < t < +\infty$.

The b-boundary consists of the plane {t=0}, but it is clear that any boundary point is regular since it is contained in the extension of the half-spacetime into the full Minkowski spacetime.

(b)- The Schwarzschild exterior solution

This is the solution of the Einstein field equations for spherically symmetric empty space. In the static region, the metric is written:

$ds^2 = -(1-2m/r)dt^2 + (1-2m/r)^{-1}dr^2 + r^2(d\theta^2 + \sin^2\theta d\phi^2)$

where $-\infty < t < +\infty$, $0 < r < \infty$, $0 \le \theta \le \pi$, $0 \le \phi \le 2\pi$

The b-boundary is the surface r = 2m, the well-known Schwarzschild horizon. But we know that Schwarzschild space-time is extensible (see, for example, Misner, Thorne and Wheeler (MTW), 1974). So, to avoid these pathologies, it is sufficient to construct the b-boundary only if the space-time is maximal, in which case there will by definition no longer be any regular boundary point.

However, the problem is not easy: the choice of extension is not trivial and the maximal analytic solution can be very complicated.

For example, Schwarzschild's outer solution can be extended analytically into Kruskal's (1960) maximal analytic solution; but an extension can also be made so as to obtain an 'inner' solution representing a static spherical star, or a spacetime containing a star evolving towards a black hole by the process of gravitational collapse.

The choice of extension therefore contains essential physical information.

Despite its interest, this problem will not be discussed further in the remainder of this paper.

*2.3. The Three Classes of Singularities*

After several attempts (Clarke, 1975a-b, King and Ellis, 1973), the following terminology will be adopted (Ellis and Schmidt, 1977).

Given a point p belonging to the b-boundary $\partial M$ of an inextensible singular spacetime M, consider the components of the curvature tensor $R_{abcd}(v)$ measured in a basis $\{E_a(v)\}$ along a curve $\gamma(v)$ ending at p, together with the components of the covariant derivatives of $R_{abcd}$ to order k: $R_{abcd\,;e1...ek}(v)$, and finally the $C^k$-*curvature scalar fields*, which are polynomial scalar fields constructed from the metric tensor $g_{ab}$, the completely antisymmetric tensor $\eta^{abcd}$, the Riemann tensor and its covariant derivatives up to order k.

The point p of $\partial M$ is said to be:

(a) $C^k$-*quasi-regular singularity* (resp, $C^{k-}$), $k \geq 0$, if the $R_{abcd\,;e1...ek}(v)$ components measured in a *parallel transported* basis along any curve ending at p are functions of class $C^0$ (resp. $C^{0-}$, i.e. locally bounded), but there is no $C^k$-extension (resp. $C^{k-}$-extension) of (M, g) en p.

(b) $C^k$-*non-scalar singularity* if there exists a curve $\gamma(v)$ such that the $R_{abcd\,;e1...ek}(v)$ components measured in a *parallel transported* basis along $\gamma(v)$ are not all functions of class $C^0$, but there exists no $C^{k-}$ irregular curvature scalar field along all curves ending in p.

(c) $C^{k-}$ *scalar singularity* if there is a curve $\gamma(v)$ ending in p and a $C^{k-}$ scalar field of curvature that is not of class $C^0$ along $\gamma(v)$.

This classification has the advantage of grouping together the various cases where the "singularity" of p refers to the behaviour of the Riemann tensor and those where it refers to the differentiability of the metric at p (cf. §1.3).

Scalar and non-scalar singularities specifically involve an irregularity of curvature in the sense that observers moving towards the singularity along certain curves can measure components of the Riemann tensor that do not have a finite limit in parallel transported reference frames. In other words, it is the curvature of space-time that prevents any extension to the point p (whereas in the case of quasi-regular singularities the obstacle is topological). These two classes can therefore be grouped together under the same heading of *curvature singularities*, also called p.p.curvature singularities ("parallelly propagated") in the notation of Hawking and Ellis (1973).

By considering which part of the Riemann tensor is irregular, we can refine the classification by defining *Ricci (or matter) singularities* if it is the components of the Ricci tensor that are irregular, and *Weyl (or conformal) singularities* if the irregularity only affects the Weyl tensor. Finally, we can distinguish cases where the non-convergence of the curvature tensor occurs because the irregular components are unbounded (*divergent* curvature singularities) or because they remain bounded but oscillate indefinitely without having a limit (*oscillatory* curvature singularities).

*2.4. Quasi-Regular Singularities*

A $C^k$-quasi-regular singularity is also said to be *$C^k$-locally extensible*, in the sense that any curve $\gamma(v)$ ending at p and satisfying certain regularity conditions (causal curve) is contained in an open U of M such that $\gamma$ can be extended beyond p in a $C^k$-local extension of (U, g) (Clarke,1973).

Here, "$C^k$-local extension" means local extension of differentiability class k, where k is the differentiability of the metric on M.

In other words, $\gamma$ is contained in a neighborhood isometric to an open of a non-singular spacetime. Physically, an observer moving along $\gamma$ towards p experiences no infinite gravitational force, because the components of the Riemann tensor measured in a tetrad transported parallel along $\gamma$ are convergent. However, there is no global extension of (M, g) into (M', g') that would send p onto an interior point of M': p is indeed an essential singularity attached to spacetime, but does not lead to singular physical effects.

The simplest example of a quasi-regular singularity is the vertex of a cone. The most famous illustration is provided by the Taub-NUT-Misner spacetime (see for example Hawking-Ellis, 1973, §5.8 and Ryan-Shepley, 1975, §8), about which we will say a few words.

In 1951, Taub discovered a solution of Einstein's equations for empty space with the property of spatial homogeneity, i.e. at each point there is a spatial hypersurface on which the metric is independent of position.

The metric of Taub space is written as:

$$ds^2 = -U^{-1}dt^2 + (2l)^2 U(d\psi + \cos\theta d\phi)^2 + (t^2+l^2)(d\theta^2 + \sin^2\theta d\phi^2)$$

where $U(t) = -1 + 2(m+l^2)(t^2+l^2)^{-1}$, m and l are positive constants, $0 \leq \psi \leq 4\pi$, $0 \leq \theta \leq \pi$, $0 \leq \phi \leq 2\pi$ (the solution has the topology $R \times S^3$, so that $\psi$, $\theta$ a,d $\phi$ are Euler coordinates on $S^3$).

This metric is singular in $t = t_\pm = m \pm (m^2 + l^2)^{1/2}$, where U = 0.

Just as the Schwarzschild solution admits an extension through the horizon r=2m, the Taub solution admits an extension beyond the surfaces $t = t_\pm$. The resulting space was described independently by Newman, Unti and Tamburino (1963), and it was Misner and Taub (1969) who highlighted the link between the two solutions: the Taub solution and the NUT solution can be "assembled" into a single manifold, the Taub-NUT-Misner space, in which there are two regions, one with the Taub metric, the other with the NUT metric, separated by thousand hypersurfaces of topology $S^3$ called Misner bridges.

The Taub-NUT-Misner space has incomplete geodesics ending in quasi-regular singularities.

A two-dimensional example given by Misner (1967) has similar properties, so it is sufficient to describe this simpler case to get an idea of the situation in Taub-NUT-Misner space-time.

Consider the two-dimensional Minkowski space (fig.3).

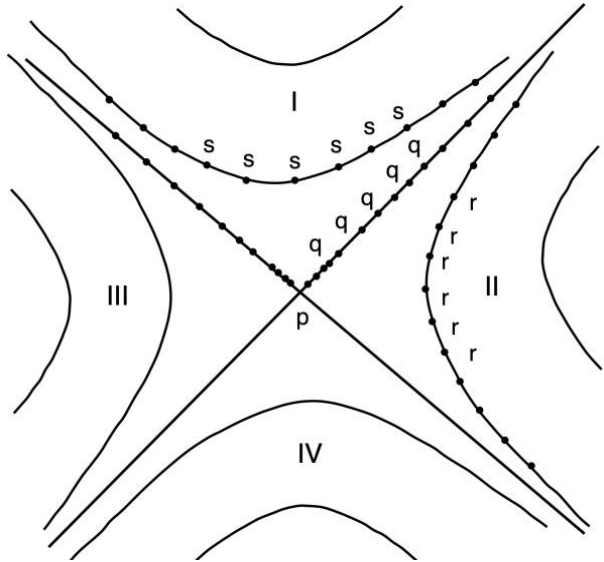

**Figure 3. Equivalence of points in 2D Minkowski space**

Under the action of the discrete subgroup G of the Lorentz group, the s points are equivalent, as are the q or r points.

By identifying the equivalent points in regions I and II, we obtain the cylinder: $ds^2 = -t^{-1}dt^2 + td\psi^2$, which has a closed null geodesic (t=0) defining a locally extensible singularity (see Figure 4).

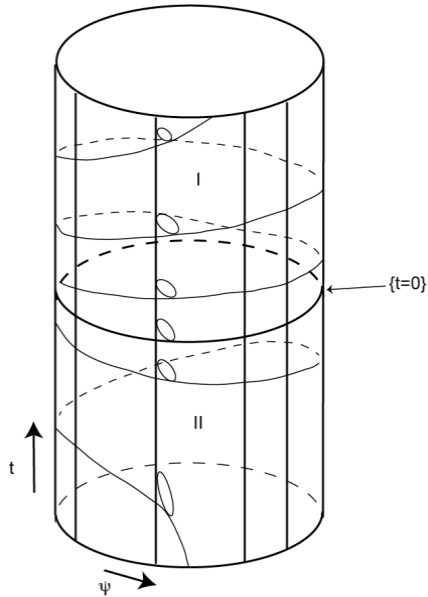

**Figure 4. Minkowski cylinder obtained by identification of points**

The vertical null geodesics are complete, but the null geodesics that spiral around the cylinder are incomplete. These geodesics can be extended locally by 'unscrewing' them, but this extension simply winds up the previously complete vertical null geodesics, rendering them incomplete.

So, there is no global extension that would allow all the geodesics to be extended simultaneously beyond t=0, but only a local extension that would allow one or other of the two families of null geodesics to be extended: t=0 is indeed a locally extendable singularity.

In fact, it is possible to construct spacetimes with quasi-regular singularities by "cutting up" and "putting back together" pieces of regular spacetime (see for example Ellis-Schmidt, 1977).

*2.5. Existence of Quasi-Regular Singularities*

The question arises as to whether these 'unnatural' singularities can develop from regular initial data.

In raising this question, we have in mind the problem of the nature of singularities in a cosmological model (i.e. a physically reasonable model capable of representing the whole of our universe during a given phase, and not just part of it). If quasi-regular singularities only occur in somewhat pathological spacetimes, they will be excluded from cosmological models.

Clarke (1975b) actually showed that, in general, any point p belonging to the boundary $\partial V$ of a globally hyperbolic manifold V and accessible by a time curve of V cannot be a quasi-regular singular point.

Recall that a manifold V is said to be globally hyperbolic if it is causal and if the causal interval between any two points of V is compact; an equivalent formulation is that there exists a surface S in V called a *partial Cauchy surface* such that any causal curve intersects it once and only once.

The theorem is valid in general, i.e. quasi-regular singularities are only accessible by time curves if the Ricci and Weyl tensors have very particular forms (the spacetime is then said to be *D-specialized*, which is physically equivalent to an empty Petrov D-type spacetime).

Clarke (1976a) improved his theorem by including null curves: a maximal globally hyperbolic spacetime (with the required differentiability class) that is not D-specialized at any point has no quasi-regular singularity accessible by a causal curve.

Thus, the Cauchy expansion of regular initial data does not in general lead to a quasi-regular singularity.

We can interpret this result by saying that the existence of a quasi-regular singularity accessible by a causal curve of a globally hyperbolic space-time must be *unstable*, since a small perturbation of the initial data may be enough to destroy the very special conditions that the curvature tensor must assume.

Another case where a quasi-regular singularity is accessible by causal curves occurs when incomplete causal curves are 'trapped' in a compact region of spacetime. A result of Hawking and Ellis (1973, proposition 8.5.2) shows that here again, the existence of the quasi-regular singularity must be unstable.

Finally, these examples suggest that quasi-regular singularities are either *primordial*, i.e. have existed since time immemorial (for a precise definition, see Clarke, 1976a) or are unpredictable *holes* developing without apparently reasonable cause from regular Cauchy data. A theorem of Clarke (1976a, Th.4) seems to confirm this hypothesis.

We shall see in the next section that the general theorems on singularities of Hawking and Penrose involve causal curves. The class of quasi-regular singularities is therefore ruled out in the context of these theorems. For this reason, we will not dwell further on this class, even though numerous examples of space-time pathologies rivalling in perversity have been given to illustrate these quasi-regular... rather singular singularities!

Finally, however, let us add that, since we can construct many relatively simple spacetimes admitting a quasi-regular singularity by cutting out particular regions of regular spacetimes and making certain identifications, we might hope to find a sub-classification of this type of singularity. In fact, it seems that this is not possible (Ellis and Schmidt, 1977, Appendices A and B).

*2.6. Non-Scalar Singularities*

The definition of non-scalar singularities suffices to show that the curve $\gamma(v)$ ending at p cannot be extended beyond p into a local $C^k$-extension.

Physically speaking, there are curves approaching arbitrarily close to the non-scalar singularity along which an observer measures perfectly regular gravitational forces, but in other cases the observer may measure irregular inertial forces (Clarke, 1976b).

A non-scalar singularity may occur in the following situation: suppose that along any incomplete curve $\gamma(v)$ ending in p, $v \in [0, v_+[$ and $\gamma(v_+) = p \in \partial M$, there exists an orthonormal basis in which the $R_{abcd\;;e_1\ldots e_k}(v)$ components remain perfectly regular along $\gamma(v)$) (clearly under these conditions p is not a scalar singularity).

The point p will be a non-scalar singularity if on a curve $\gamma(v)$ ending at p the basis $\{Y_i\}$ is connected to a basis $\{E_a\}$ transported in parallel along $\gamma(v)$ by an irregular Lorentz transformation $\Lambda^{a}{}_i$, i.e. if $Y_i(v) = \Lambda^{a}{}_i E_a(v)$, at least one of the functions $\Lambda^{a}{}_i(v)$ is not of class $C^k$ (or $C^{k-}$) on $[0, v_+]$.

In this case, the curvature tensor is perfectly regular on the approach to p, in the sense that its components are regular when $\{Y_i\}$ is used as a basis; but the curvature tensor measured in a parallel transported basis has an irregular behaviour.

In Clarke's (1975b) notation, a non-scalar singularity that occurs in such circumstances is called an *intermediate singularity*.

*Example 1: Plane-wave spacetime*

These are solutions of the field equations for empty space, homeomorphic to $R^4$, with metric :

$ds^2 = dx^2 + dy^2 + 2dudv + 2H(x,y,u)du^2$

where the coordinates (x,y,v) vary from $-\infty$ to $+\infty$, with

$H(x,y,u) = \frac{1}{2}A(u)[(x^2-y^2)\cos\theta(u) - 2xy\sin\theta(u)]$ where $A(u)$, $\theta(u)$ are arbitrary functions of class $C^1$ defined on an open I of R, determining the amplitude and polarisation of the wave.

These spaces are invariant under a five-parameter group of isometries $G_5$ which is simply transitive on null surfaces $\{u = cste\}$ (Ehlers and Kundt, 1962).

Consider the orthonormal reference frame $\{E_a\}$ defined by:

$E_1 = \frac{\partial}{\partial x}$, $E_2 = \frac{\partial}{\partial y}$, $E_3 = \frac{1}{\sqrt{2}}\left(\frac{\partial}{\partial u} + (1-H)\frac{\partial}{\partial v}\right)$, $E_4 = \frac{1}{\sqrt{2}}\left(\frac{\partial}{\partial u} - (1-H)\frac{\partial}{\partial v}\right)$

The components of the curvature tensor are determined by the components of the Ricci tensor and the Weyl tensor $C_{abcd}$. The latter is in turn determined by the symmetrical 3-tensors of zero trace $E_{ab}$ ("electric" part of the Weyl tensor) and $H_{ab}$ ("magnetic" part) (see for example Ellis, 1971). In the $\{E_a\}$, we find:

$E_{ab} \equiv C_{a4b4} = \begin{pmatrix} \alpha(u) & \beta(u) & 0 \\ \beta(u) & -\alpha(u) & 0 \\ 0 & 0 & 0 \end{pmatrix}$, $H_{ab} \equiv C_{a4b4}^{*} = \frac{1}{2}\eta_{a4}{}^{de}C_{de4b} = \begin{pmatrix} \beta(u) & -\alpha(u) & 0 \\ -\alpha(u) & -\beta(u) & 0 \\ 0 & 0 & 0 \end{pmatrix}$

and $R_{ab}= 0$ \hfill (2.1)

where $\alpha(u) = A(u)\cos\theta(u), \quad \beta(u) = A(u)\sin\theta(u)$ \hfill (2.2)

The basis $\{E_a\}$ is parallelly transported along the timelike geodesic $\gamma(s)$ of equation ($x=y=0$, $u=-v=\frac{1}{\sqrt{2}}s$), of which $E_4$ is the tangent vector (H=0 on this curve). This geodesic is equivalent to any other time geodesic because of the invariance of the metric under the $G_5$ group of isometries.

We now define the field of orthonormal tetrads $\{Y_i\}$ by:

$Y_1 = \cos\frac{1}{2}\theta(u)E_1 + \sin\frac{1}{2}\theta(u)\ E_2$

$Y_2 = \cos\frac{1}{2}\theta(u)E_2 - \sin\frac{1}{2}\theta(u)\ E_1$

$Y_3 = ch\ \xi(u)E_3 + sh\ \xi(u)E_4$

$Y_4 = ch\ \xi(u)E_4 + sh\ \xi(u)E_3$

where $\xi(u)$ is a function of arbitrary class $C^1$.

In this basis, the curvature tensor takes the form (2.1) with:

$$\alpha(u) = A(u)\exp 2\xi(u), \quad \beta(u) = 0 \tag{2.3}$$

Plane-wave spacetimes are geodesically complete if A(u) and q(u) are functions of class $C^1$ defined for $-\infty < u < +\infty$, in other words if $I \equiv R$. On the other hand, if either of the functions A(u), q(u) diverges for a finite value $u_+$, a non-scalar singularity appears on $\gamma(s)$, since according to (2.1) and (2.2) the curvature tensor diverges in $u_+$ in the parallelly transported basis $\{E_a\}$. But if we use the $\{Y_i\}$ basis with, for example, $\xi(u) = -\frac{1}{2}$ Log$(1+A^2(u))$, (2.3) leads to $\alpha(u) = A(u)/(1+A^2(u))$, $\beta(u)=0$, so that the components of the curvature tensor are perfectly regular in this basis (they are continuous and bounded by $\pm\frac{1}{2}$).

Under these conditions, all $C^0$-scalar invariants formed from $R_{abcd}$, $g_{ab}$ and $\eta^{abcd}$ will behave regularly along $\gamma(s)$.

*Example 2:* *''Tilted '' Spatially homogeneous models (Ellis and King, 1974)*

These are spatially homogeneous spacetimes with perfect fluids, where the 4-velocity of the fluid is not parallel to the 4-vector unit normal to the homogeneous hypersurfaces. Some of these models have a non-scalar singularity (picturesquely called a "whimper", which could be translated as "crybaby"), a simple idea of which can be obtained by examining a two-dimensional model given by Ellis and King (I974).

Consider the two-dimensional Minkowski space and the action of the Lorentz group around a point p (fig.3).

The Lorentz transformations act in surfaces at a constant distance from p and leave p fixed.

Let's draw on this space a line of universes of the cosmological fluid not passing through p, and crossing from region I to region II. Let's apply the Lorentz transformations to this line of universes. We obtain a "stack" of fluid universe lines on the null line that separates regions I and II from regions III and IV (fig.5).

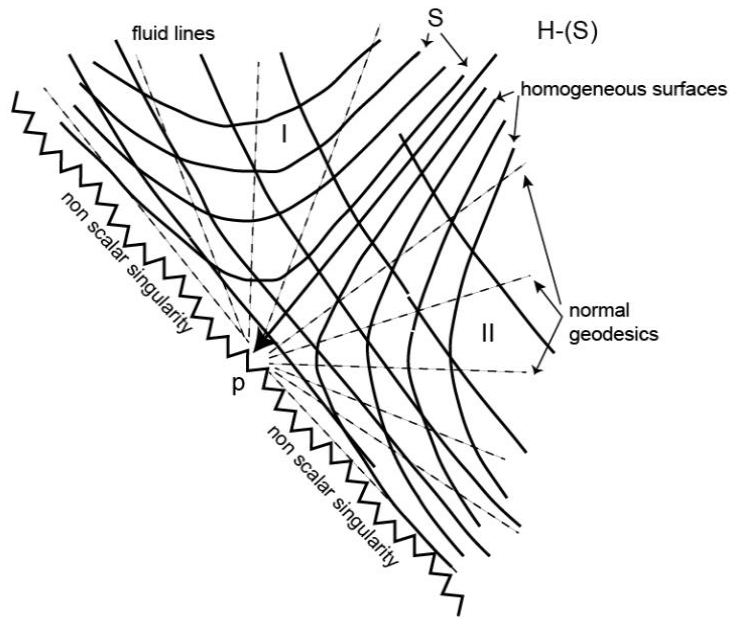

**Figure 5. Construction of a "whimper" singularity**

This null line is singular and corresponds to an intermediate singularity. Density and pressure are finite at any point arbitrarly close to the singularity; in fact, the limiting value of these quantities is precisely that which they take on the null surface H⁻(S), which is a homogeneous surface on which density and pressure are constant and perfectly defined.

Region I is spatially homogeneous (the homogeneity surfaces are spatial), while region II is stationary (the homogeneity surfaces are temporal).

Let's consider an observer moving towards the singularity along a geodesic orthogonal to the homogeneity surfaces in region I.

He measures a matter density $\bar{w} = T_{ab}n^a n^b$ ($n^a$ being the normal vector). Since homogeneous surfaces become spatial in region I after being temporal in region II, the normal vector $n^a$ becomes null on H⁻(S), so the scalar product $u_a n^a$ diverges on homogeneity surfaces tending towards H⁻(S) and p ($u^a$ is the fluid velocity). The stress-energy tensor of a perfect fluid has the form:

$T_{ab} = (w+p)u_a u_b + p g_{ab}$, we see that $\bar{w} = (w+p)(u_a n^a)^2 - p$ becomes infinite as we approach p. So the observer moving towards p measures an infinite density of matter.

This situation is an exact analogue of what happens in four dimensions. The curvature tensor measured in an orthonormal frame of reference along an orthogonal geodesic ending at p diverges if the frame of reference is transported parallel; on the other hand, it is perfectly regular in another orthonormal frame of reference along the curve (for example a tetrad with the 4-velocity of the fluid as the timelike vector).

The null homogeneous surface H⁻(S) is a *prediction horizon* or *Cauchy horizon*, in the sense that complete initial data on a spatially homogeneous surface S of the region I completely determines the evolution of the field between S and H⁻(S), but not beyond.

An example of such a model is the "tilted" Bianchi type V universe found by Farnsworth (1967), whose Penrose diagram has the shape shown in figure 6.

This example leads to the following question: is a non-scalar singularity always associated with a Cauchy horizon?

The answer is negative (Ellis and Schmidt, 1977), although certain solutions which are globally hyperbolic (i.e. without a Cauchy horizon) and have non-scalar singularities provide a counter-example.

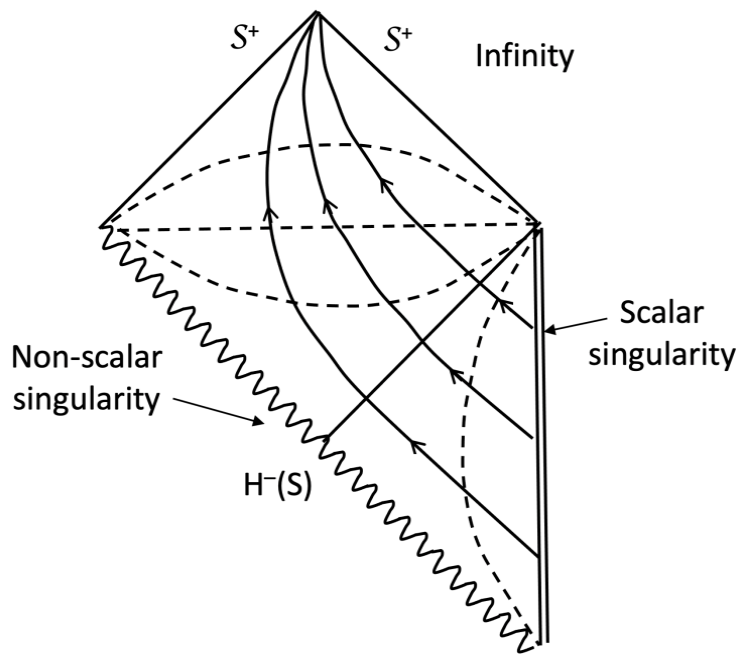

**Figure 6. Penrose diagram of a tilted Bianchi type V universe**

*2.7. Existence of Non-Scalar Singularities*

Let us list some very recent results relating to the occurrence of non-scalar singularities

(a) Empty spaces of Petrov types III or N can only possess $C^0$-non-scalar singularities (or locally extensible singularities), this being due to the fact that there is no $C^0$-non-zero scalar invariant.

(b) There are vacuum solutions with $C^0$- non-scalar singularities, in which the Petrov type is not constant in an open neighborhood of the singular point p. Consider for example a plane-wave solution in which the function A(u) is of class $C^-$ on $(0,\infty)$, cancels on any interval $(1/2^n, 1/2^{n+1/2})$, where n is an integer, and is non-zero on any interval $(1/2^{n+1/2}, 1/2^{n+1})$ of maximum amplitude $n^2$.

When $n \to \infty$, A(u) oscillates indefinitely with Petrov types N and O. From (2.3) and (2.5), it is clear that we obtain a $C^0$-non scalar singularity.

(c) Empty spaces of Petrov types D, I and II can have $C^0$-non-scalar singularities but also $C^0$-scalar singularities (unlike the other types, see above).

(d) All $C^0$-non-scalar singularities (whether Weyl, Ricci or mixed singularities) can be described as due to an irregular Lorentz transformation between a "good" tetrad and a parallel transported tetrad (Siklos, 1976).

(e) In general, non-scalar singularities are both Weyl and Ricci singularities (Siklos, 1976).

(f) A study of the possible behaviour of the curvature tensor along a curve approaching a non-scalar singularity has been carried out by Ellis and Schmidt (1977); one of the conclusions is that $C^0$-non-scalar singularities can occur in spaces with positive definite metrics, but $C^{0-}$-non-scalar singularities cannot.

To conclude this section, let us mention that so far there is no theorem relating to the "instability" of non-scalar singularities, in the sense that if a spatially homogeneous model possessing a non-scalar singularity is subjected to perturbations, the non-scalar singularity would transform into a scalar singularity. Indeed, if a photon were sent through the Cauchy horizon $H^-(S)$, there would be a spectral shift towards infinite blue as it approached the non-scalar singularity (Ellis and King, 1974). The photon would then arrive with infinite energy, and the effect of the perturbation would be to transform the non-scalar singularity into a scalar singularity. Some work by Belinskii, Khalatnikov and Lifshitz (1972) and King (1975) suggests that non-scalar singularities are indeed unstable because they would be linked to conditions of strong symmetry in space-time, so that in general cosmological models (without symmetry) such a situation would not occur.

*2.8. Scalar Singularities*

In the notation of Hawking et Ellis (1973), scalar singularities are *s.p.curvature singularities* ("scalar polynomial").

Ellis and King (1974) showed that in a perfect fluid spacetime whose equation of state p = p(w) is such that the inequalities: w+3p ≥ 0, w+p > 0, 0 ≤ dp/dw ≤ 1 are verified, then a "curvature singularity" appears on a given curve if and only if a scalar singularity appears there. The "curvature singularity" referred to here is defined according to the terminology of certain articles (Clarke, 1975b; Ellis and King, 1974) such that the components of the curvature tensor measured in any orthonormal reference frame along the curve are divergent.

This type of singularity is both the best known and the closest to the intuitive physical conception of a singularity, as the following examples show.

*(1) Friedmann-Lemaître Universe*

These are solutions of Einstein's equations that are spatially homogeneous and isotropic, with a perfect fluid. The metric can be written as

$$ds^2 = -dt^2 + R^2(t)d\Sigma^2$$

where $d\Sigma^2$ is the metric of a 3-space of constant curvature k = +1, 0 or -1. The fluid moves at 4-velocity $u^a = \delta^a_0$.

We show that if w+3p > 0 and w+p > 0, there is a scalar singularity at the beginning of each expansion phase: w $\to\infty$ when t$\to$0 and the Ricci curvature invariant $R^{ab}R_{ab}$, diverges (see §5.3).

*(2) Schwarzschild inner solution*

The well-known metric is written :

$$ds^2 = -(1-2m/r)dt^2 + (1-2m/r)^{-1}dr^2 + r^2(\sin^2\theta\, d\phi^2 + d\theta^2)$$

with $0 \leq \phi \leq 2\pi$, $-\infty < t < \infty$, $0 < r < 2m$.

r=0 is a scalar singularity, because the conformal curvature invariant $R^{abcd}R_{abcd}$ becomes infinite there.

*(3) Orthogonal spatially homogeneous cosmologies*

i.e. with a perfect fluid whose universe lines are orthogonal to the homogeneous hypersurfaces) or class A tilted cosmologies (King and Ellis, 1973). These models will be studied in detail in a second article.

*(4) Stationary solutions with cylindrical symmetry and incoherent matter (King, 1974).*

These spaces admit a three-parameter abelian group of isometries acting transitively on temporal hypersurfaces; the temporal Killing vector is tangent to these surfaces. Some solutions, such as that of Maitra (1966), are geodesically complete, while others have a Weyl singularity (the scalar $E^{ab}g_{ab}$ becomes infinite, where $E^{ab}$ is the "electric" part of the Weyl tensor), or an oscillatory Ricci singularity (the Ricci scalar polynomial $R^{ab}R_{ab}$ is of the form $2\sin^2(1/r) + f(r)$, where f is monotonic, and therefore has no limit when r tends to zero).

Presumably, the existence of scalar singularities is stable under small perturbations. Here again there is no definitive theorem, but there are strong presumptions (Eardley, Liang and Sachs, 1972; Lifshitz and Khalatnikov, 1970).

Note that proving the existence of a scalar singularity is not always trivial. Despite Cauchy data that would formally lead to a scalar singularity, it is mathematically possible that the Cauchy development is precisely not the one expected; spacetime may, for example, 'end' before the predicted singularity occurs. We need to ensure that the predictions (or retrodictions) made on the basis of formally adequate Cauchy data are not distorted by the spontaneous appearance of 'unexpected' singularities (for example, a conformal singularity with infinite matter density). To achieve this, we postulate that spacetime is maximal and has no "*holes*", i.e. that for any spatial surface S with no boundary, the dependence domain D(S) is such that there is no isometry $\Phi$: D(S)$\to$N in another spacetime N for which D($\Phi$(S))$\neq \Phi$(D(S)) (Clarke, 1976).

For example, the universal cover of the Minkowski space amputated from the 2-plane (t=0, x=0) is maximal but has a hole; indeed, D({t=- 1}) is "holed" by the singularity at t=x=0, and its image under the natural application $\Phi$ which sends it into the Minkowski space is a proper subset of D($\Phi$({t=-1})), which is the entire space.

Using D instead of D$^+$ makes the definition symmetric between prediction and backtracking, which avoids having to decide which time arrow to choose.

We won't dwell on the properties of scalar singularities here, as we'll be studying their various structures in detail in a further article.

## 3. Occurrence of Singularities

*3.1. Theorems*

We will briefly deal with the part relating to the famous theorems on the occurrence of singularities in Cosmology, theorems which, with the discovery of the cosmic background radiation at 2.7°K, were at the origin of a great revival of interest in singularities among cosmologists. As this subject has been remarkably and exhaustively dealt with in the classic work by Hawking and Ellis (1973), we shall confine ourselves to listing these theorems without demonstration and discussing their conditions and fields of application.

We have already pointed out in §1.1 that some physicists in the 1960s regarded singularities in General Relativity simply as the result of symmetries imposed on space-time. The most representative authors of this school, Lifshitz and Khalatnikov, published a series of works showing that certain classes of solutions with a "physical" singularity (infinite matter density or pressure) did not contain the number of arbitrary functions required to specify a general (i.e. symmetry-free) solution of Einstein's equations (Lifshitz and Khalatnikov, 1963). The two authors then hypothesized that the Cauchy data giving rise to such singularities formed a subset of zero measure in the set of all Cauchy data, and therefore did not occur in the real universe.

We have already seen that this hypothesis has in fact been invalidated by the proof of general theorems on the occurrence of singularities.

The first theorem not involving a symmetry assumption was formulated by Penrose (1965) and demonstrates the occurrence of a singularity in a star undergoing gravitational collapse below its Schwarzschild radius.

However, the Schwarzschild surface is only defined for exact spherical symmetry. Penrose therefore introduced the more general concept of a *closed trapped surface* **S,** which is a compact, edgeless 2-space surface (whose topology is usually $S^2$), such that the two families of geodesics (incoming and outgoing) orthogonal to **S** each converge on **S**.

The rigorous statement is :

<u>Theorem 3.1</u> The spacetime (M,g) cannot be n-g-complete if all the following conditions are satisfied:

(1) $R_{ab}K^aK^b \geq 0$ for any null vector $K^a$

(2) there exists a noncompact Cauchy surface in M

(3) there exists a closed trapped surface **S** in M.

Let us briefly discuss the conditions of the theorem.

Condition (1) holds provided that the energy density is positive for any observer.

Condition (3) holds at least in some region of spacetime (see Hawking and Ellis, 1973, §9).

The relative "weakness" of the theorem lies in condition (2) which imposes the existence of a Cauchy surface, so that the theorem does not predict the existence of singularities in physically realistic cosmological models. In fact, it shows that a gravitational collapse leads either to a singularity (more precisely, an n-g-incompleteness) or to a Cauchy horizon - either way, the prediction is broken.

A theorem by Hawking and Penrose (1970) overcomes this difficulty:

<u>Theorem 3.2</u> (M, g) is c-g-complete if:

(1) $R_{ab}K^aK^b \geq 0$ for any null vector $K^a$

(2) the generic condition is satisfied

(3) M satisfies the causality condition

(4) at least one of the following three conditions is satisfied:

**(4i.)** there exists an achronal compact set without edge

(4ii) there is a closed trapped surface

(4iii) there exists a point p such that on any null geodesic from p the divergence θ becomes negative.

This theorem is more general than the first. On the one hand, condition (1) corresponds to the weak energy condition, which involves not only zero vectors as in Theorem 3.1, but also time vectors; on the other hand, the existence of a closed occlusive surface is now only one of three possible conditions.

It can be shown that (4i) is satisfied in a spatially closed solution; (4iii) is satisfied if the reconverging light cone is our own light cone from the past (observation of radiation at 3°K). The genericity condition (2) is written as

$K^a K^b K_{[c} R_{d]ab[e} K_{f]} \neq 0$ for any causal geodesic with tangent unit vector $K^a$; it is reasonable in any physically realistic solution (Hawking and Ellis, 1973, p.101).

However, Theorem 3.2 does not specify whether the singularities occur in the past or the future. If (4i) is verified, the singularity is in the future; if (4ii) is verified, nothing can be said; if (4iii) is verified, the singularity is in the past.

A theorem of Hawking (1967) solves this question:

<u>Theorem 3.3</u> There exists an incomplete causal geodesic in the past passing through a given point p of spacetime M if:

(1) $R_{ab} K^a K^b \geq 0$ for any causal vector $K^a$

(2) M satisfies the causality condition

(3) At point p there exists a temporal unit vector of the past W and a constant b>0 such that if V is the unit vector tangent to the temporal geodesics of the past passing through p, on each of these geodesics the expansion $\theta \equiv V^a{}_{;a}$ becomes less than $-3c/b$ (where $c = -W^a W_a$) in the distance interval [0, b/c] measured from p along these geodesics.

Condition (3) simply means that the past-time geodesics from p reconverge within a compact region located in the past of p. This situation is illustrated in Figure 7.

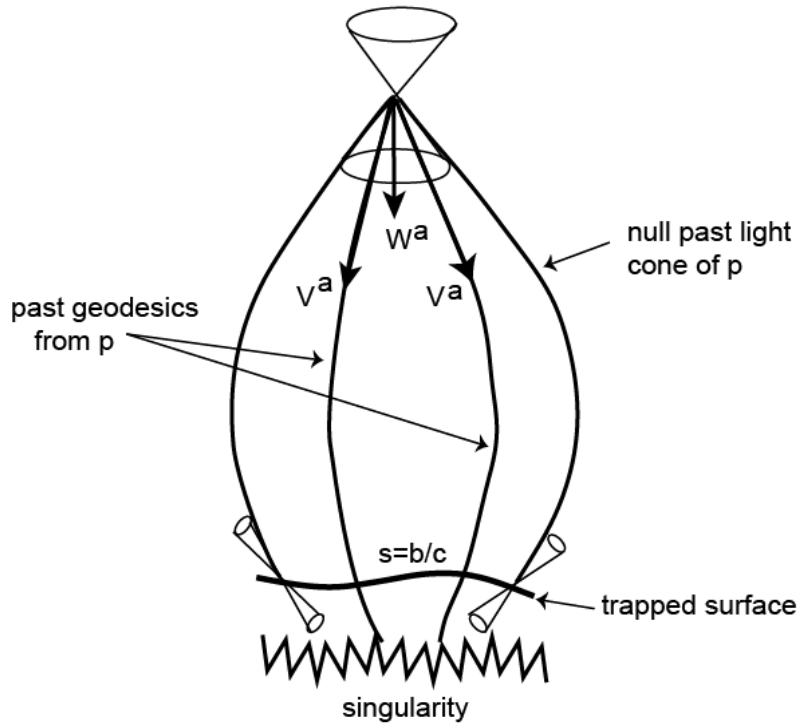

**Figure 7. Reconvergence of past-time geodesics within a trapped surface**

Theorems 3.2 and 3.3 are the most useful because their assumptions can be shown to be satisfied in a large number of solutions of the physically realistic Einstein equations.

Finally, these theorems involve the condition of causality, and one could envisage a violation of causality rather than an incompleteness of space-time. A theorem by Hawking (1967) shows that this is not the case:

*Theorem 3.4* Spacetime is t-g-incomplete if:
   (1) $R_{ab}K^aK^b \geq 0$ for any causal vector $K^a$
(2) there exists a compact 3-edgeless space surface **S**
(3) the unit normals to **S** are everywhere divergent (or everywhere convergent) on **S**.

is the weak energy condition; (2) refers to the closure of the universe; (3) is interpreted as saying that the universe is expanding (or contracting).

Geroch (1966) also contributed to these questions with a theorem close in substance to theorem 3.4.

*3.2. Conclusion*

As we have seen, these extremely important theorems prove, within the framework of General Relativity and under reasonable physical hypotheses, the existence of a singularity in the past of the universe or a singularity in the future as far as the gravitational collapse of stars is concerned.

In fact, they simply tell us that space-time has incomplete causal geodesics. These incomplete geodesics define singular points on the b-boundary ∂M, but we know nothing about the physical nature of these singularities.

In an attempt to elucidate this fundamental question, several directions are open to the researcher, that remain relevant today. Among them:

(a) The Penrose-Hawking singularity theorems prove the inevitability of spacetime singularities in general relativity under classical and rather reasonable conditions, but they say very little about their nature ((Senovilla, 2012). "Weak" singularities in which all the scalar quantities remain finite can occur in the so-called tilted Bianchi cosmologies. However, such cosmological models are likely not generic. Belinskii, Khalatnikov and Lifshitz (1970, 1982) [BKL] have conjectured that for a generic inhomogeneous relativistic cosmology, the approach to the spacelike singularity into the past is vacuum dominated and oscillatory, obeying the so-called BKL or mixmaster dynamics – for a synthesis, see Belinskii and Henneaux (2018). Numerical simulations have been used, e.g. by Berger (2002) and Garfinkle (2017), to verify the BKL dynamics in special classes of spacetimes. Rigorous mathematical results on the dynamical behaviour of Bianchi type VIII and IX cosmological models have also been presented in Brehm (2016). The interest in considering generic solutions is that the behaviour of a general solution (without symmetry) in the vicinity of a singularity can be considered qualitatively similar to the behaviour of these particular generic solutions. In a second part of the present review on spacetime singularities, we shall discuss the oscillatory approach to singularities in Bianchi-type cosmological models.

(b) We can try to weave the link between the singularities predicted by General Relativity and those studied in other branches of Physics. In this respect, René Thom's theory of catastrophes (1969) could provide some interesting results, but as far as I know this way has not been developed. However, while catastrophe theory remains largely unexplored in this context, dynamical systems techniques, like phase space and stability analysis, have provided powerful tools for studying the evolution and classification of singularities, especially in relation to BKL dynamics, see e.g. Bahamonde et al. (2018), Roy, Chanda & Paul (2024).

(c) We could also consider the "brutal" method of numerically integrating Einstein's equations on a computer. Full 3D simulations of Einstein's field equations became possible only with the advent of supercomputers and robust numerical relativity formalisms in the 1990s and early 2000s, see e. g. Baumgarte & Shapiro (1998). They have become essential for exploring dynamical spacetime evolution, shedding light on black hole interiors, gravitational collapse, and phenomena like critical behavior – e.g., Choptuik (1993). More recent work addresses mass inflation, weak singularities, and Cauchy horizon stability, e.g. Hod and Piran (1998).

(d) In order to sketch a bridge to current work in the study of singularities, it is useful to recall that much of the mystery of singularities hinges on a unified theory of quantum gravity, which could provide a deeper understanding of these exotic objects. The term "soft singularities" has become more prevalent in recent years, particularly in the context of black hole physics and quantum gravity. It is used to describe singularities where the breakdown of spacetime geometry is "milder" or less extreme than what we typically expect from classical singularities, like those associated with black holes or the Big Bang. They arise in various contexts, including string theory and fuzzball models (Mathur, 2005), quantum cosmology and quantum bounce (Ashtekar and Singh, 2011), black hole evaporation (Bañados and Cano, 2015), soft hair models (Hawking and Perry, 2016), quantum gravity models that smooth the infinite curvatures of traditional singularities (Giddings and Karczmarek, 2009). While the exact nature and mathematical formulation of soft singularities remain an open area of research, they are part of an effort to understand and resolve the infinities predicted by classical general relativity through quantum mechanical effects.

*Acknowledgement*
The author thanks the anonymous referees for their helpful comments that improved the quality of the manuscript.

## References

1. Ashtekar A, Singh P. (2011) "Loop quantum cosmology: A status report" *Class. Quant. Grav. 28(21), 213001*.
2. Bahamonde S., Boehmer C.G., Carloni S., Copeland E.J., Fang W., Tamanini N. (2018) "Dynamical systems applied to cosmology : dark energy and modified gravity", *Physics Reports Vol.775-777 pp.1-122*
3. Bañados M., Cano P.A. (2015) "The final state of a black hole: Thermal or singular?", *J. High Energy Physics 2015(8), 1-10*.
4. Baumgarte T., Shapiro S. (1998) "On the numerical integration of Einstein's field equations" *Phys. Rev. D 59, 024007*
5. Belinskii V.A., Henneaux M. (2018) *The Cosmological Singularity*, Cambridge University Press. [arXiv:1404.3864]
6. Belinskii V. A., Khalatnikov I. M., Lifschitz E. M. (1970) "Oscillatory Approach to a Singular Point in the Relativistic Cosmology", *Adv. Phys. 19, 525.*; (1982) "A general solution of the Einstein equations with a time singularity", *Adv. Phys., 31 6, pp. 639-667.*
7. Belinskii V.A., Lifshitz E.M., Khalatnikov I.M. (1972) "Construction of a General Cosmological Solution of the Einstein Equations with a time Singularity" *Sov.Phys.JETP, 35, 838*
8. Berger B.K. (2002) "Numerical Approaches to Spacetime Singularities" *Living Rev. Rel. 5 1*
9. Brehm B. (2016) "Bianchi VIII and IX vacuum cosmologies: Almost every solution forms particle horizons and converges to the Mixmaster attractor" *[arXiv:1606.08058]*
10. Brans C., Dicke R.H. (1961) "Mach's Principle and a Relativistic Theory of Gravitation" *Phys. Rev. 124, 925*
11. Choptuik M.W. (1993) "Universality and scaling in gravitational collapse of a massless scalar field", *Phys. Rev. Lett. 70, 9*
12. Choquet-Bruhat Y. (1968) "Espace-temps einsteiniens généraux et chocs gravitationneIs" *Ann.Inst.H.Poincaré, 8, 327*
13. Christodolou D. (1984) "Violation of cosmic censorship in the gravitational collapse of a dust cloud", *Comm. Math. Phys. 93(2): 171-195* ; (1999) "The instability of naked singularities in the gravitational collapse of a scalar field", *Ann. Math.*, vol. 149, nº 1, pp. 183–217.
14. Clarke C.J.S. (1973) "Local extensions in Singular Space-Times" *Commun.Math.Phys. 32, 205*
15. Clarke C.J.S. (1975a) "The Classification of Singularities" *Gen.Rel.Grav. 6, 35*
16. Clarke C.J.S. (1975b) "Singularities in Globally Hyperbolic Space-Times" *Commun.Math.Phys. 41, 65*
17. Clarke C.J.S. (1976a) "Space-Time Singularities" *Commun.Math.Phys.49, 17*
18. Crowther K., De Haro S. (2022) "Four Attitudes Towards Singularities in the Search for a Theory of Quantum" In *A. Vassallo, The Foundations of Spacetime Physics: Philosophical Perspectives. New York, NY: Routledge. pp. 223-250*.

19. Curiel, E. (2019) "Singularities and Black Hole", in *The Stanford Encyclopedia of Philosophy*, ed. E.Zalta, https://plato.stanford.edu/
20. Dafermos M. (2014) "Black holes without spacelike singularities", *Commun. Math. Phys., Volume 332 (2014) no. 2, pp. 729-757*
21. Eardley D.M., Liang K., Sachs K. (1972) "Velocity-dominated singularities in irrotational dust cosmologies" *J.Math.Phys.* 13, 99
22. Ehlers J., Kundt W. (1962) "Exact solutions of the gravitational field equations", in *Gravitation: an introduction to current reasarch*, ed.Witten& Wiley, New York
23. Ellis G.F.R.(1971): "Relativistic Cosmology", in *General Relativity & Cosmology*, ed.R. Sachs, XLVII Enrico Fermi School, Acad. Press.
24. Ellis G.F.R., King A.R. (1974) "Was the big-bang a whimper?" *Commun. Math. Phys. 38, 119*
25. Ellis G.F.R., Schmidt B.C. (1977) "Singular Spacetimes", *Gen. Rel. Grav. 8, 915–953*
26. Farnsworth D.L. (1967) "Some New General Relativistic Dust Metrics possessing Isometries", *J.Math.Phys. 8, 2315*
27. Friedrich H. (1974) "Construction and properties of space-time b-boundaries" *Gen. Rel. Grav. 5, 681*
28. Garfinkel D. (2017), "Numerical relativity beyond astrophysics", *Rep. Progr. Phys. 80 016901*
29. Geroch R.P. (1966) "Singularities in closed universes", *Phys. Rev. Lett. 17, 445*
30. Geroch R.P. (1968a): "What is a singularity in General Relativity?" *Ann.Phys. 48, 526*
31. Geroch R.P. (1968b): "Local characterization of singularities in General Relativity", *J.Math.Phys. 9, 450*
32. Geroch R.P., Kronheimer E., Penrose R. (1972) "Ideal points in space-time" *Proc. Roy. Soc. Lond. A 327,545*
33. Giddings S.B., Karczmarek J. (2009) "A new approach to black holes and cosmology in the AdS/CFT context", *Phys. Rev. D 80(6), 065011.*
34. Hawking S.W.(1966) "The occurrence of singularities in Cosmology" *Proc.Roy.Soc.Lond. A 294, 511; 295, 490*
35. Hawking S.W. (1967) "The occurrence of singularities in Cosmology III. Causality and singularities" *Proc.Roy.Soc.London A300, 187*
36. Hawking S.W. (1969) "On the rotation of the universe" *MNRAS, 142, 129*
37. Hawking S.W. (1974) "Black hole explosions?" *Nature, 248, 30*
38. Hawking S.W. (1975) "Particle creation by black holes" *Comm.Math.Phys. 43, 199*
39. Hawking S.W. (1976) "Breakdown of Predictability in Gravitational Collapse", *Phys.Rev.D 14, 2460*
40. Hawking S.W., Ellis G.F.R. (1973) *The large scale structure of space-time,* Cambridge University Press
41. Hawking S.W., Penrose R. (1970) "The singularities of gravitational collapse and Cosmology", *Proc.Roy.Soc.Lond. A314, 529*
42. Hawking S.W., Perry, M.J.,Strominger A. (2016) "Soft Hair on Black Holes", *Phys. Rev. Lett. 116(23), 231301.*
43. Hehl F.W., von der Heyde P., Kerlick G.F. & Nester, J.M. (1976) "General relativity with spin and torsion: Foundations and Prospects", *Reviews of Modern Physics, 48(3), 393-416*
44. Hicks N.J. (1965) *Notes on Differential Geometry,* D. VanNostrand, Princeton
45. Hod S., Piran T. (1998) "Mass-Inflation in Dynamical Gravitational Collapse of a Charged Scalar-Field", *Phys.Rev. Lett. 81, 1554-1557.*
46. Joshi P.S. (2014) "Spacetime Singularities ", in *Springer Handbook of Spacetime, eds.A. Ashtekar and V. Petkov,* p.409-436, Springer
47. King A.R. (1974) "New Types of Singularity in General Relativity" *Commun.Math.Phys. 38, 157*
48. King A.R. (1975) "Instability of Intermediate singularities in, general relativity" *Phys.Rev.D 11, 763*
49. King A.R., Ellis G.F.R. (1973) "Tilted homogeneous cosmological models". *Commun.Math.Phys. 31, 209*
50. Kobayashi S., Nomizu K. (1963) *Foundations of Differential Geometry* (Interscience)
51. Kruskal M.P. (1960) "Maximal extension of Schwarzschild metric", *Phys.Rev. 119, 1743*
52. Kundt W. (1963) "Note on the completeness of space-times", *Zs. f.Phys. 172, 488*
53. Landsman K. (2021) "Singularities, black holes and cosmic censorship: A tribute to Roger Penrose [*https://arxiv.org/abs/2101.02687*]
54. Lifshitz E.M., Khalatnikov I.M. (1963) "Investigations in relativistic cosmology" *Adv.Phys. 12, 18*
55. Lifshitz E.M., Khalatnikov I. M. (1970) "Oscillatory approach to singular point in the open cosmological model" *JETP Lett. 11, 123*
56. Maitra, S.C. (1966) "Stationary Dust-Filled Cosmological Solution with Λ=0 and without Closed Timelike Lines", *J. Math. Phys., 7, 1025.*
57. Mathur S.D. (2005) "String theory and Black hole Fuzzballs : a brief review", *International Journal of Modern Physics A 20(27), .2009-2028.*
58. Misner C.W. (1967a) "Taub-NUT space as a counterexample to almost anything", in *Relativity & Cosmology, ed.J. Ehlers, Lectures in Applied Math. vol.8, p. 160* (Am.Math.Soc., Providence)
59. Misner C.W., Taub A.H. (1969): "A singularity-free empty universe" *Sov.Phys. JETP, 28, 122*
60. Misner C.W., Thorne K.S., Wheeler J.A. (1974) *Gravitation,* Freeman (San Francisco)

61. Newman E.T., Tamburino L., Unti T.J. (1963): "Empty space generalization of the Schwarzschild metric", *J.Math.Phys.* <u>4</u>*, 915*
62. Penrose R. (1965) "Gravitational Collapse and space-time singularities" *Phys. Rev. Lett. 14, 57*
63. Penrose R. (1969). "Gravitational Collapse: the Role of General Relativity" *Nuovo Cimento, Rivista Serie.* <u>1</u>*, 252.*
64. Roy B.C., Chanda A., Paul B.C. (2024) "Dynamical stability and phase space analysis of an Emergent Universe with non-interacting and interacting fluids" [https://arxiv.org/abs/2401.00782]
65. Ryan M.P., Shepley L.C. (1975): *Homogeneous Relativistic Cosmologies,* Princeton University Press (New Jersey)
66. Schmidt B.G. (1971) "A new definition of singular points in General Relativity", *Gen.Rel.Grav. 1, 269*
67. Schmidt B.G. (1972) "Local completeness of the b-boundary", *Commun. Math.Phys.* <u>29</u>*, 49*
68. Seifert H.J. (1975) "The causal structure of singularities", in *Conference on Differential Geometrical Methods in Mathematical Physics*, 539-565, Hamburg.
69. Senovilla J. M. M. (2012), "Singularity theorems in general relativity: achievements and open questions", Chapter 15 of *Einstein and the Changing Worldviews of Physics*, eds. C. Lehner, J. Renn and M. Schemmel, Einstein Studies 12 (Birkhauser, 2012)
70. Siklos S. (1976) "Two completely singularity-free nut spacetimes", *Phys. Lett. A,* <u>59(3)</u>*, 173-174 (1976)*
71. Sobolev S.L. (1963) "Applications of Functional Analysis to Physics" vol.7, *Translations of Math. Monographs (Am.Math.Soc.)*
72. Steenrod N. (1951) *The Topology of Fibre Bundles*, Princeton Univ.Press
73. Thom R. (1969) *Stabilité Structurelle et Morphogénèse*, Benjamin (N-York)
74. Van de Moortel M. (2025) "The Strong Cosmic Censorship Conjecture", *Comptes Rendus de l'Académie des Sciences. Mécanique, Volume 353, pp.415-454*